# Architected Dual-Network Solvent-free Adhesives for Stretchable Fabrics

*Gabriela Moreira Lana, Cornelia Meissner, Siddhant Iyer, Xin Hu, Perin Jhaveri, Skylar Tibbits, Alfred J. Crosby\**

G. Moreira Lana, C. Meissner, S. Iyer, X. Hu, A. J. Crosby

Department of Polymer Science and Engineering, University of Massachusetts Amherst, Conte Center for Polymer Research, 120 Governors Drive, Amherst, MA 01003, USA

E-mail: acrosby@umass.edu

P. Jhaveri

Department of plastics engineering, University of Massachusetts Lowell, 1 University Ave, Lowell, MA 01854, USA

S. Tibbits

School of Architecture + Planning, MIT Massachusetts Institute of Technology, 77 Massachusetts Avenue, Cambridge, MA 02139, USA

Natural systems, such as tendons and spider silk, demonstrate how the combination of strength and stretchability can be effectively achieved by integrating stiff and flexible network structures. Inspired by these systems, we developed a novel, solvent-free dual-network adhesive based on a self-assembling ABA triblock copolymer, poly(methyl methacrylate)-poly(n-butyl acrylate)-poly(methyl methacrylate) (PMMA-b-PnBA-b-PMMA), designed for applications requiring both high strength and stretchability. The triblock copolymer forms a physically crosslinked network through microdomains of PMMA end-blocks that provide structural integrity, while the PnBA mid-block forms a soft, stretchable matrix. To further enhance mechanical performance, a second poly(n-butyl acrylate) (PnBA) network is polymerized in situ, locking the PMMA microdomains in place and creating a load-bearing system. By varying the crosslinking density of the secondary network, we tailor the adhesive's mechanical properties (Young's modulus: 0.17 – 1.18 MPa) to suit different substrates, creating a mechanically transparent seam. The resulting dual-network system combines different strategies to achieve high strength and stretchability, with adhesive performance comparable to industrial methods such as sewing, particularly in bonding neoprene fabric composites and sealing the joints. Our solvent-free approach also eliminates the need for lengthy solvent evaporation steps, offering an eco-friendly and more efficient alternative for flexible adhesive applications in fields such as soft robotics, flexible electronics, and sports apparel.

## 1. Introduction

Joining soft, flexible materials, such as fabrics and multilayered stretchable composites, presents unique challenges due to their surface topography, porosity, and the need to maintain mechanical properties under large deformations.[1] Conventional solutions, such as seam tapes, often use hot-melt adhesives, which can be too stiff and lead to stress concentrations[2] and poor mechanical integration with the substrate under deformation.[3,4] Similarly, solvent-based adhesives rely on the use of multiple solvents to wet and penetrate the substrate, but they introduce lengthy evaporation times and environmental and health concerns. Without solvents, poor penetration can lead to delamination or stress concentrations at the joint, ultimately causing poor adhesion and failure.[5,6] Sewing is another common approach, but it introduces failure points at the needle holes[7] and requires complex machinery, limiting durability.[8,9] Thus, there is a need for innovative adhesive systems that offer strong, flexible, and durable bonds without relying on solvents or stitching.

To address these challenges, we looked to nature for inspiration, focusing on biological systems that serve structural roles and act as natural adhesives binding tissues and structures of different properties together. Biological materials such as tendons and spider silk illustrate how gradients in mechanical properties, combined with different networks and hierarchical microstructures, can bind components with varying stiffness, ensuring effective load transfer without compromising flexibility. Tendons serve as transition structures between stiff bone and soft muscle, distributing mechanical forces while maintaining both strength and compliance. Their primary structural component is a collagen network, which provides strength and mechanical resistance, while the dense connective tissue surrounding the collagen fibers allows for the tendon's stretchability. [10,11] Another example is dragline spider silk, which features a hierarchical structure of stiff nanocrystalline β-sheets embedded within a softer amorphous protein matrix, as represented in **Figure S1**, enabling it to combine robustness and elasticity.[12] Inspired by natural systems, we designed our adhesive with dual functionality, achieving both strength and stretchability to match the mechanical properties of the soft, fabric-based adherend.

To this end, we utilized a dual-network approach[13–15], integrating both physical and chemical crosslinks,[16–18] combined with self-assembly into high-order architectures.[19,20]

Our system consists of an ABA triblock copolymer with poly(methyl methacrylate) (PMMA) end blocks that self-assemble into rigid domains, and a poly(n-butyl acrylate) (PnBA) mid-block that forms a soft, elastic matrix.[21–23] A second network is formed through in situ photo-polymerization of a butyl acrylate (BA) monomer, which serves as a mid-block selective solvent during seam bonding, as represented in **Figure 1b**. This polymerization locks the PMMA domains in place, creating a load-bearing system capable of large-strain, large force, and reversible deformations.

Perhaps most importantly, by adjusting the crosslinking density[24] of the secondary network, we can tailor the adhesive's mechanical properties to suit, and even match, those of different adherends. This allows for the fabrication of "mechanically transparent seams", meaning that the bonded region replicates the mechanical response of the adherends, behaving as if it were a continuous material. The material's tunability allows for the application on challenging materials which require flexibility and strength under large deformations. Furthermore, unlike traditional adhesives for fabric composites, our system eliminates the need for solvents, offering an eco-friendly and efficient solution that can be directly applied to the adherends.

To demonstrate the effectiveness of our adhesive, we selected a formulation that matches the mechanical properties of neoprene fabric and applied it directly as a joining material without solvents, achieving good penetration into the fabric layer (**Figure 1a**), strong adhesion, stretchability, and barrier performance. Neoprene fabric, a multi-layered composite, is utilized across various industries, including sports apparel, medical devices, fashion, industrial gaskets and seals, marine equipment, military gear, and soundproofing panels. In all of these applications, and many others where textiles and soft, stretchable materials are used, our adhesive offers an innovative, eco-friendly, and efficient alternative by eliminating solvents while delivering excellent flexibility, strength, and durability.

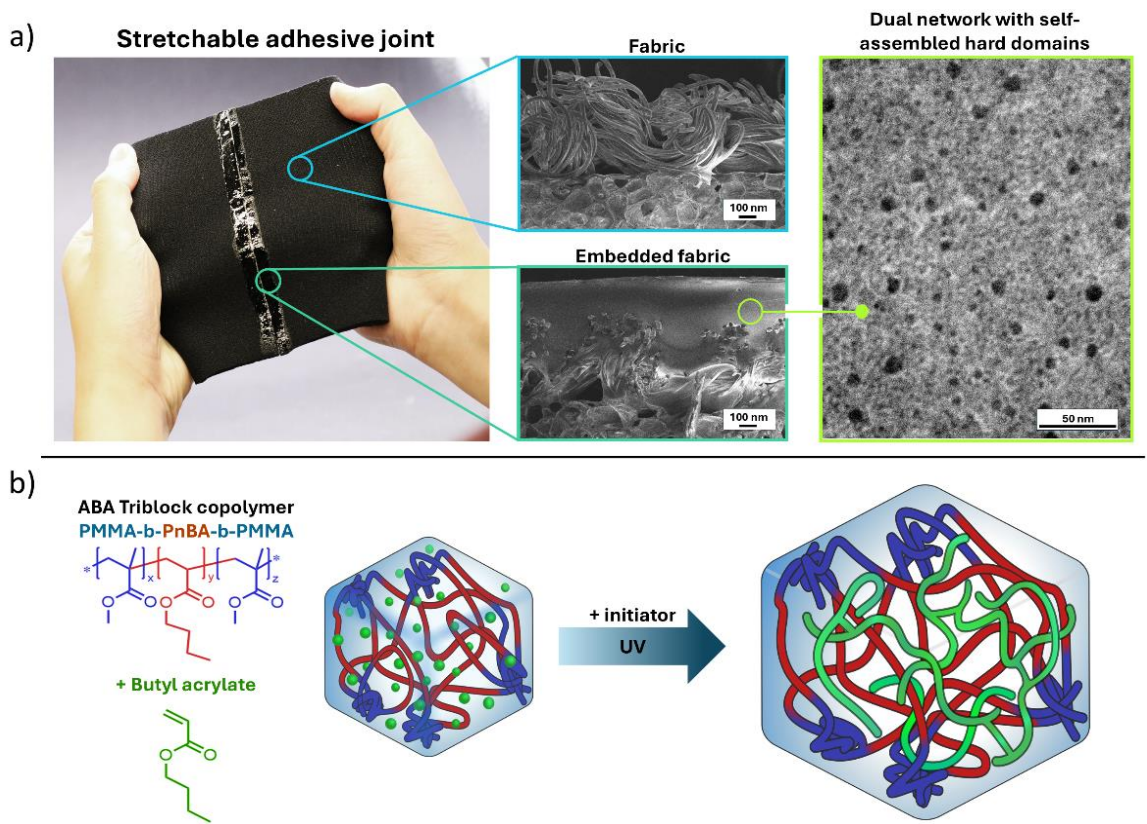

*Figure 1: Dual-network adhesive strategy for stretchable and soft materials*. *(a) Adhesive bonding of neoprene fabric substrates while maintaining stretchability. Scanning electron microscopy (SEM) images show the outer fabric layer of the neoprene composite, and the region where the adhesive was applied shows the embedded fabric. Transmission electron microscopy (TEM) image of the dual network reveal self-assembled hard domains that contribute for excellent mechanical properties. b) Schematic representation of the ABA triblock copolymer Poly(methyl methacrylate)-block-poly(n-butyl acrylate)-block-poly(methyl methacrylate) (PMMA-b-PnBA-b-PMMA) that self-assembles when dissolved in the butyl acrylate monomer. Photopolymerization under UV forms a secondary network of poly(n-butyl-acrylate) (PnBA), locking the domains in place and enhancing mechanical properties of the material.*

## 2. Results and discussion

The adhesive formulation was prepared by dissolving the ABA triblock copolymer PMMA-b-PnBA-b-PMMA pellets in the n-butyl acrylate (BA) monomer, which served as a midblock-selective solvent. Addition of photo-initiator (phenylbis(2,4,6-trimethylbenzoyl)phosphine oxide, Irgacure 819) and, for some samples, a crosslinker (ethylene glycol dimethacrylate, EGDMA) enable the *in situ* polymerization of BA into a PnBA network. Samples were polymerized under UV irradiation (7.5-9 mW/cm²) for 5, 15 or 30 minutes. The initial mixture consisted of 60 wt% BA and 40 wt% triblock copolymer. Triblock concentrations above 40 wt% were not characterized for mechanical and adhesion properties due to a significant increase in solution viscosity, which impaired handling and application.

To verify the BA polymerization within the triblock structure, we conducted gel permeation chromatography (GPC) and Fourier transform infrared spectroscopy (FTIR) on samples without any crosslinker, but with varying initiator content (1 mol% and 2 mol% of the number of vinyl groups) and UV exposure times (5, 15, and 30 minutes). Specific weight fraction of the components is reported in **Table 1**.

**Figure 2a** shows the GPC traces of samples with different initiator contents and different UV exposure times, compared to the neat triblock sample. The response was normalized by the signal of the triblock, which was used at the same concentration for all formulations. The triblock was detected at ~21 minutes indicating a molecular weight of ~66,000 g/mol against PMMA standards. The broader signal observed for the formulations, overlapping with the triblock, is attributed to the PnBA polymerized *in situ*. For reference, polymerization of BA without triblock (using the same polymerization times and initiator concentrations) yielded the polymers analyzed in **Figure S2a**, confirming that the broad peaks at ~20 min correspond to PnBA. The average of molecular weight values, $M_w$ observed in Figure S2a for the neat PnBA was 109,811 g/mol ± 26,318 g/mol. Considering the entanglement molecular weight for PnBA of ~28,000 g/mol,[25,26] this verified the formation of a physically entangled secondary PnBA network in the system.

**Figure S2b** includes the GPC traces of the formulation samples exposed to UV for 5 min. These samples were not further analyzed due to the low PnBA yield, as indicated by the estimated relative content of PnBA shown in **Figure S2c**.

**Figure 2b** depicts the FTIR spectra of the liquid adhesive mixture (before photopolymerization) and the respective films (after photopolymerization) for samples with 1 mol% and 2 mol% initiator, exposed to UV for 15 or 30 min. The depletion of the C=C stretching modes at 1619 and 1636 cm$^{-1}$ in the samples after UV exposure indicate monomer consumption and confirms the completion of the photopolymerization reaction.[27]

Subsequently, to assess the effects of crosslinker content, all crosslinked samples were tested with 2 mol% initiator. A similar effect was observed in FTIR for samples with fixed initiator content and varying amount of crosslinker (2.5 mol%, 5 mol%, and 10 mol% as function of the BA monomer concentration), as shown in **Figure S2d**, indicating quantitative reaction for all samples cured for 15 or 30 min. A detailed description of weight fraction of each component in the final formulation is presented in Table 1 in the materials and methods section.

To confirm the absence of unreacted monomers, polymerized samples were stored at 40 °C for 24 h and mass changes were recorded. Gravimetric measurements showed a mass loss of 0.2 to 0.35 wt% after 24 h, indicating full conversion. The mass of the samples remained stable for another 3 days at room temperature (~23 °C) following the 24-hour treatment, confirming the effectiveness of the photopolymerization reaction. Additionally, thermogravimetric analysis (TGA) was performed immediately after UV for the sample containing 2 mol% initiator and 5 mol% crosslinker, exposed to UV for 15 min, as shown in **Figure S3**. The TGA with a heating rate of 2 °C/min revealed a 0.44% mass loss up to 145 °C, which corresponds to the boiling point of the monomer BA. A more significant mass loss, observed by a change in slope, takes place above around 165 °C due to thermal degradation of the polymer, in agreement with data reported in the literature[28]. The initial minor mass loss corroborates the minimal loss of volatile and unreacted compounds after photopolymerization of the secondary network, observed in the gravimetric analysis through post-treatment at 40 °C; TGA confirms stability of the adhesive even at higher temperatures (up to ~165 °C).

Subsequently, we conducted an elution test to confirm the formation of a network upon addition of the crosslinker EGDMA to the initial mixture of triblock, BA, and initiator. A polymerized sample (5 mol% crosslinker, 15 min UV) was immersed in tetrahydrofuran (THF) for ~30 min (Inset of **Figure 2c**), and the eluted material was analyzed using NMR spectroscopy. The analysis revealed the absence of BA in the eluted material, as shown by the comparison with the pure BA spectrum in **Figure 2c**, which shows the vinyl hydrogen peaks. When comparing the spectrum of the eluted material with that of the original triblock (**Figure 2d**), there was almost complete overlap. Analyzing the NMR spectrum we conclude that ~75 wt% of the leached material is triblock and ~25 wt% can be attributed to the presence of poly(butyl acrylate) oligomers that were not integrated into the crosslinked PnBA network. The washed crosslinked sample was then dried at room temperature for 48 h, resulting in a final weight of approximately 48 % of the initial sample weight. This mass reduction also suggests the removal of the triblock (40 %) and a small fraction of non-crosslinked PnBA.

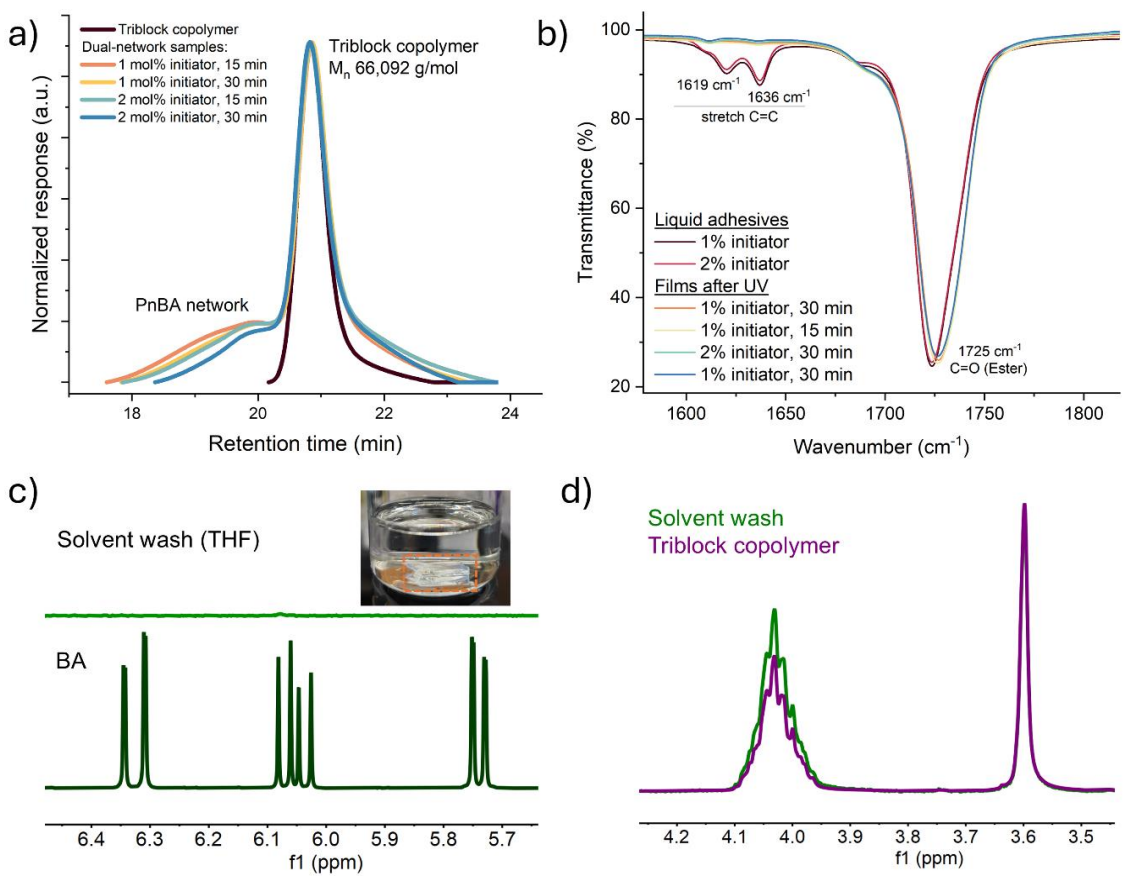

*Figure 2: Analysis of the dual network formation through PnBA polymerization within the triblock copolymer structure. a) GPC traces of polymerized samples without crosslinker, with different initiator content (1 % or 2 %) and UV exposure time (15 or 30 min), in comparison to the neat triblock. b) FTIR spectra of samples with varying initiator before and after UV, indicating complete reaction of the double bonds after photopolymerization. Similar outcome was observed for samples with varying crosslinker contents and is presented in Figure S2. c) NMR analysis from THF elution test from a sample with 5 mol% crosslinker: the spectrum indicates the absence of butyl acrylate monomers. d) NMR spectrum of the eluted matter shows the presence of the triblock and a small fraction of PnBA. This indicates the dissolution of the triblock and retention of the major PnBA crosslinked network.*

Previous studies of PMMA-b-PnBA-b-PMMA and other symmetric triblock systems dissolved in midblock selective solvents report microphase separation into spherical or cylindrical structures. For example, in a PMMA-b-PnBA-b-PMMA system in 2-ethyl-1-hexanol (midblock-block-selective solvent at low temperatures), the transition between these morphologies occurs for a PMMA weight fraction ranging from 0.07 to 0.24.[29] Comparably, the triblock used in this study has two PMMA segments with weight fraction of 0.15 (0.3 total PMMA weight fraction). However, the morphology within systems can vary, as the interaction with the solvent is the driving force for phase separation, and is determined by polymer concentration and processing history, such as annealing. Neat triblock samples with varying concentration of end and middle blocks, and triblock samples with added homopolymer show the same phase transition, indicating that, to some extent, we can consider the total PnBA content in our modified samples as a shift in composition while keeping the Flory-Huggins segment interaction parameter and chain length similar.[30] This corresponds to a horizontal shift in a phase diagram of the Flory-Huggins segment interaction parameter as a function of system composition.

To evaluate the resultant morphology following *in situ* polymerization of the PnBA network, we performed small-angle X-ray scattering (SAXS) and transmission electron microscopy (TEM). **Figure 3a** shows the 1D scattering profiles for the neat triblock copolymer and samples with varying crosslinker contents, 0 mol%, 5 mol%, and 10 mol%, respectively. The scattering profiles were shifted vertically for clarity. The characteristic peak at a scattering vector Q=0.032 Å arises from the periodicity of the microphase domains, reflecting the average d spacing ($d = 2\pi/Q$) of ~19.6 nm between scattering centers for all samples.[31] Comparing the 1D scattering profile of the control triblock and the dual-network samples, the primary peak remains unchanged, yet the dual-network samples present a broader peak, indicating a broader distribution. The scattering pattern for higher Q values, which provides information on the form factor of the domains, does not reveal a higher order of organization.

TEM analysis of neat triblock copolymer and different dual-network samples reveals distinct morphology depending on the crosslinker content. In the TEM micrographs, the darker regions correspond to the PMMA blocks, which were stained with phosphotungstic acid (PTA) for better contrast.

**Figure S4** shows a TEM image of the neat triblock copolymer, which presents spherical domains of PMMA. According to self-consistent field theory (SCFT) predictions by Matsen and Thompson,[32] ABA symmetric triblock copolymers in the spherical phase form approximately 75% to 80% bridges. This has been corroborated by later studies using simulations[33,34] and experiments such as mechanical measurements,[21] rheology studies[35] and comparison of dieletric relaxation behavior.[36] These reports supported the idea that bridges dominate over loops in ABA triblock systems, particularly at higher polymer concentrations due to the entropic penalty of having both end blocks reside in the same micelle.[22] As polymer concentration decreases, an increased average spacing between micelles results in bridges needing to stretch further, making loop formation more favorable.

These observations are now widely accepted, and the role of bridged domains in symmetric triblock copolymers has been leveraged in various systems to enhance viscoelasticity, mechanical strength, and other physical properties.[37–40] Understanding that elastically active chains increase with the number of bridges, contributing to enhanced modulus and mechanical toughness, we selected the initial concentration of the triblock copolymer in the BA monomer to be as high as possible (40 wt% triblock dissolved in BA), while maintaining a viscosity that allows for the direct application as an adhesive or seam.

**Figure 3c** shows TEM images of the dual-network samples with varying crosslinker concentrations. In the sample without crosslinker (0 mol%), the spherical PMMA domains are relatively small and uniformly distributed. The inset in red shows an example of microdomain contours used to calculate object diameters and mean nearest neighbor distances. For this sample, the mean object diameter was ~7.0nm, and the mean spacing center-to-center between microdomains was ~14.6 nm.

At 5 mol% crosslinker content in **Figure 3c**, the microdomains maintain their spherical morphology, but some larger domains can be observed at lower magnification (upper image). This suggests that crosslinking of the PnBA homopolymer leads to some degree of aggregation or domain coarsening. The calculated mean object diameter for the higher magnification (bottom image) was ~6.4 nm and mean nearest neighbor distance was ~17.7 nm (excluding larger domains observed at lower magnification). The sample presents a broader size distribution, with domains reaching up to 18 nm, compared to the sample without crosslinker, as shown in the plot in **Figure 3b**.

Finally, with 10 mol% crosslinker, there is a notable change in the morphology. Fewer spherical domains are present, and some regions (highlighted in orange) resemble a cylindrical phase distribution. These different regions are dispersed in a matrix that does not have a distinct phase behavior, potentially composed mostly of the crosslinked PnBA phase in a random, amorphous state. Manual measurements of the spherical domains indicate diameters of around 18 nm.

The spacings of samples with 0 mol% and 5 mol% crosslinker (14 and 17 nm) as recorded by TEM are in the same range as measured using SAXS.

We hypothesize that the sample with 0 mol% crosslinker exhibits a homogeneously distributed spherical morphology because the non-crosslinked PnBA phase can ocupy the PnBA-block domain of the triblock copolymer. This allows for better mixing and a more uniform structure. When comparing the system without crosslinker to the one with 5 mol% crosslinker, we observe a shift in self-assembly behavior, leading to the formation of both large and small populations of spherical PMMA domains. In the low crosslinked system, the PnBA phase has reduced mobility, which may limit the penetration into the PnBA block domains. As the PnBA component becomes less miscible due to the increase in the crosslinking density, the midblock chains are restricted from stretching freely, causing the microdomains to separate further or aggregate, as seen in the sample with 10 mol% crosslinker.

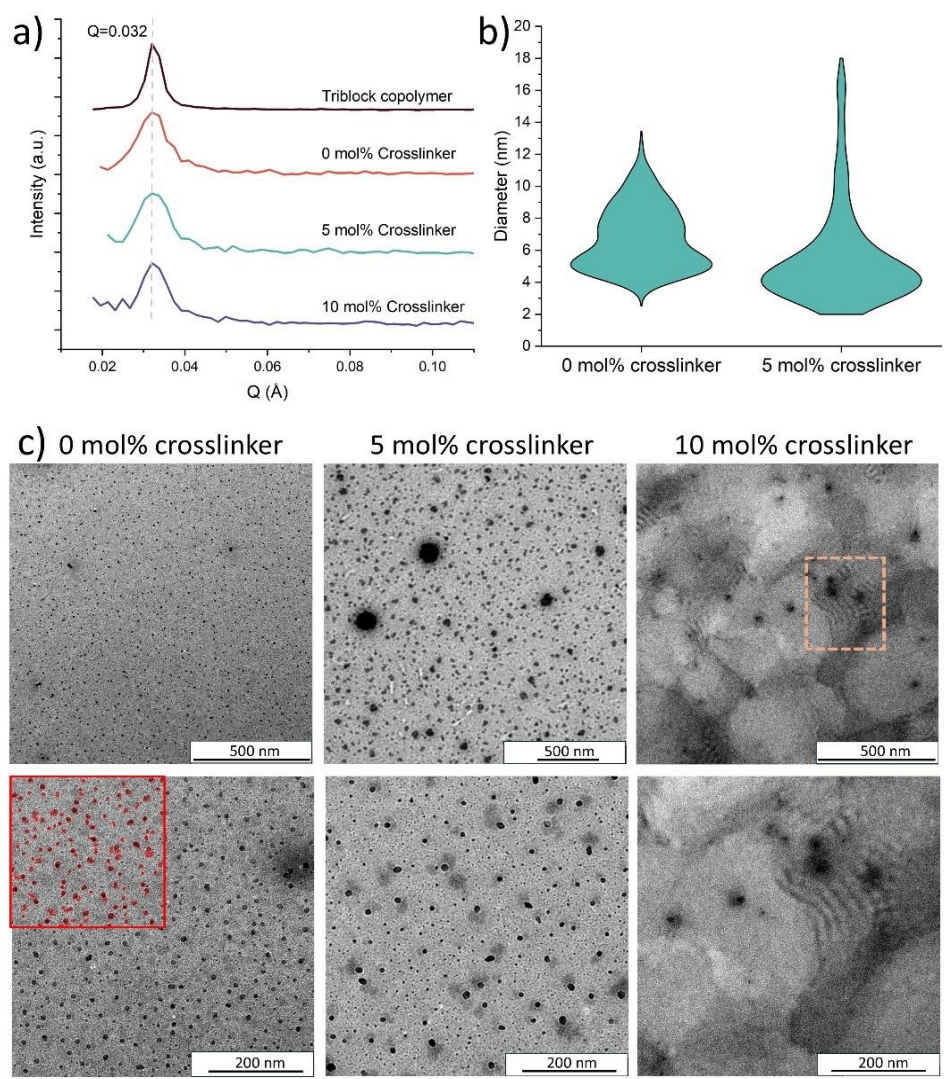

*Figure 3: Triblock copolymer self-assembly and morphology investigation. a) SAXS profiles for the triblock compared to adhesive formulations with 0 mol%, 5 mol% and 10 mol% crosslinker. b) Spherical domain diameter distribution for samples with 0 mol% and 5 mol% crosslinker, based on TEM images. c) TEM images of adhesive formulation samples, overview (top) and close-up (bottom). Samples with 0 mol% crosslinker (inset in red of example of contours detected for the microdomains, plotted in b), with 5 mol% crosslinker, and with 10 mol% crosslinker (dashed box indicates close-up region), respectively.*

**Figure 4** shows the mechanical characterization of UV-cured cast films of approximately 1 mm thickness across various formulations. **Figure 4a and 4b** show frequency sweep DMA results for samples without and with the crosslinker, respectively. In **Figure 4a**, the neat triblock copolymer as well as all the samples with the secondary PnBA network in absence of crosslinker behave as an elastic solid, with E'>>E", verifying our previous claim that the PMMA domains observed in TEM are predominantly bridged[41,42], therefore dominating the mechanical response contribution of the triblock copolymer network.[21,29] Considering that the second network of PnBA is physically entangled,[25] both the storage (E') and loss (E") moduli for all formulations (E'~ 0.2 to 0.6 MPa and E" ~ 0.06 to 0.13 MPa at 1 Hz) are lower than the neat triblock (E' ~ 2.12 and E" ~ 0.48 MPa at 1 Hz). This can be explained by the fact that the PnBA dilutes the PMMA content, compared to the neat triblock, reducing the number of hard domains that act as physical, dynamic crosslinks to dissipate energy, reducing the overall mechanical response and creating a softer material.[43]

Upon introducing different concentrations of crosslinker while keeping the UV irradiation at a constant 15 min, we observe an increased response in both storage and loss moduli (at 1 Hz, E' increases from 0.2 to 2.8 MPa as we increase the crosslinker from 0 to 10 mol%, while E" increases from 0.07 to 0.84 MPa) (**Figure 4b**). A more detailed plot of DMA with varying crosslinker content for both 15 and 30 min UV exposure times is presented in **Figure S4.** The sample with the highest amount of crosslinker, 10 mol%, exhibits storage moduli comparable to that of the neat triblock, along with higher loss moduli, indicating more dissipation in the stiffer network.[44,45] The strain hardening observed for all samples in DMA measurements is consistent with previously reported behavior for ABA triblock copolymers and systems combining triblocks with their respective mid-block homopolymer. This effect is attributed to the initial extension of the mid-block bridging chains that occur before the PMMA glassy domains are disrupted, which creates a higher effective modulus at higher deformation. [23,46]

This response is particularly interesting for adhesive applications in textile-based composites, which exhibit anisotropic, non-linear mechanical behavior. The strain-hardening in textiles at high deformations arise from architecture-dependent mechanisms, beginning with thread tensioning followed by reinforcement through interlocking designs.[47,48] This highlights the importance of molecular design of our adhesive to match the mechanical properties of the intended adherend and effectively control strain hardening, in particular focusing on the balance between crosslinking density and the role of soft mid-block of the triblock copolymer.

The trends observed in DMA results align with tensile test data represented in **Figure 4c** where increasing crosslinker content correlates to higher failure stress but lower strain at failure for both UV times of 15 and 30 min. It is important to note that, for the tensile tests of the bulk material, the samples failed at the grip, so the maximum values reported in **Figure 4c** may not fully represent the failure performance of the samples. Lastly, in **Figure 4d**, we present the Tangent moduli obtained from the stress-strain curves at low (initial 10%) and high (final 10%) strains, as a function of crosslinker content. Samples without crosslinker present a modulus of 0.17 MPa at low strain and 0.34 MPa at high strain; with 2.5 mol% crosslinker, the modulus varies from 0.28 MPa to 0.48 MPa; with 5 mol% crosslinker, the modulus increases from 0.51

MPa to 1.07 MPa; and samples with 10 mol% crosslinker vary from 1.18 MPa to 2.35 MPa. Note that the Tangent modulus values obtained at the low strain, linear regime, approach the Young's modulus of the material. The strain-hardening behavior is observed by a notable increase in modulus as strain progresses.[49] Specifically, the modulus within each formulation increases from low strain to high strain by 92.85% to 109.91%.

These results show a trade-off between the stiffness and stretchability of the dual network, by adding varying crosslinker contents and shifting from a physically to a chemically crosslinked network. The additional structural integrity obtained by the highly crosslinked samples leads to higher stiffness but also reduces energy dissipation and the ability to undergo larger deformations before failure. Considering this wide range of mechanical responses, it is possible to adjust the formulation to match the substrate's mechanical properties.

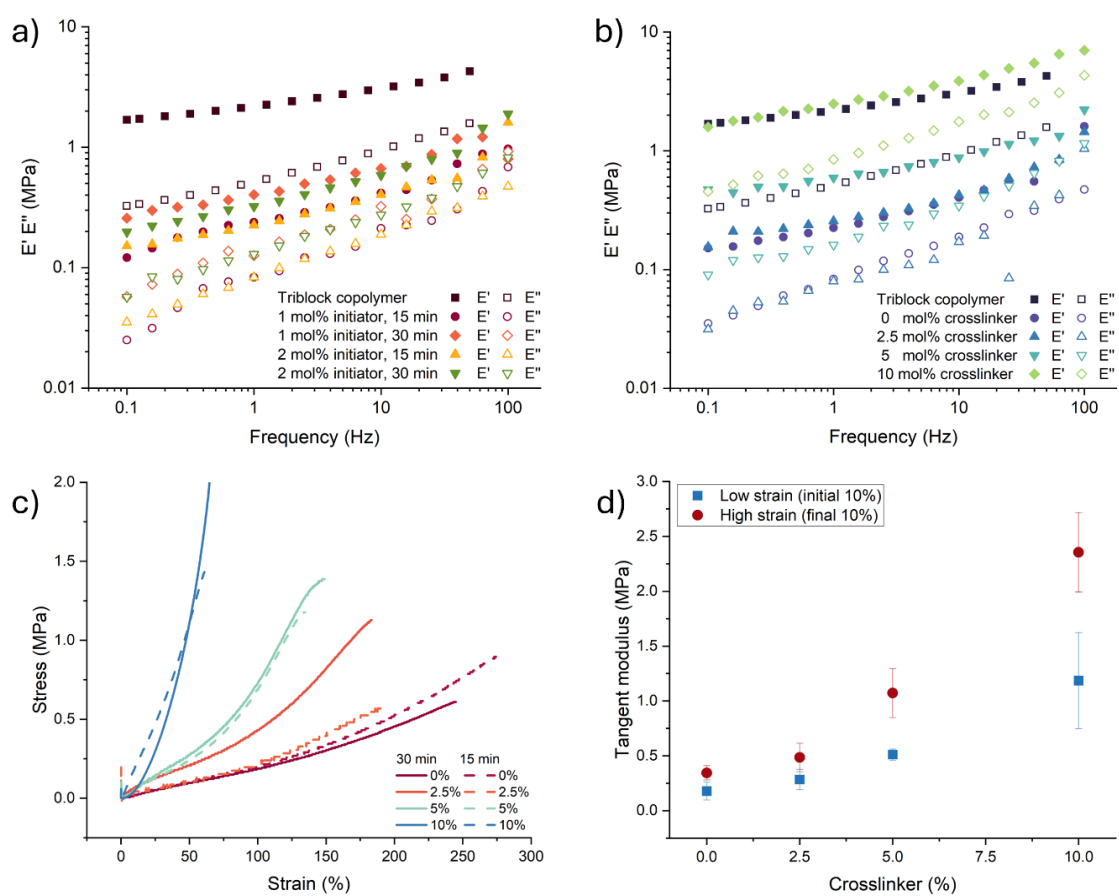

*Figure 4: Mechanical characterization of bulk films under varying conditions. a) DMA frequency sweep results for samples without crosslinker, showing the effect of varying initiator content (1 mol% and 2 mol%) and UV exposure time (15 and 30 min), compared to the neat triblock. b) DMA frequency sweep results for samples with different crosslinker contents (0 mol%, 2.5 mol%, 5 mol%, and 10 mol%), compared to the neat triblock. All presented samples were exposed to UV for 15 min (comparison with 30 min polymerization time is provided in Figure S5). c) Representative stress-strain curves from tensile tests on the bulk films,*

*illustrating the mechanical response as function of crosslinker content and UV exposure time. d) Tangent modulus calculated from the tensile tests at low strain (initial 10%) and high strain (final 10%), plotted against varying crosslinker contents. Tangent modulus approach the Young's modulus for the materials in the low strain, linear regime.*

Given the tunability of its mechanical properties, we evaluate next the adhesive's applicability for neoprene fabric – a complex, multilayered composite material consisting of outer fabric layers around neoprene foam, which presents unique challenges for bonding technologies. In its application, neoprene fabric must offer flexibility, stretchability, durability, and resilience to environmental factors. However, its surface irregularities and porosity, combined with the need for the adhesive to withstand large deformations without delamination, make achieving strong adhesion particularly challenging.

**Figure 5a** shows the stress-strain curves of the formulation materials as a bulk film with varying crosslinker content, and UV exposure 15 min, alongside the mechanical response of a commercial neoprene fabric sample up to ~100 % strain. Notably, the formulation with 5 mol% crosslinker exhibited a mechanical behavior closely matching that of the neoprene substrate and was selected as a candidate for further testing as an adhesive.

In addition to material properties, the geometry of seam construction plays a crucial role in determining the performance and durability of joints between adherends. By exploring different bonding methods, such as applying the adhesive material only in the center between the adherents (glue), or applying it as a layer on the outer region of the joint (seam), we assessed how these factors impact the overall strength and stretchability of neoprene fabric joints, compared to sewn samples.

**Figure 5b** shows the failure stress and strain at failure for the neoprene substrate and joints bonded using different methods: the formulation with 5 mol% crosslinker used as glue between two pieces of neoprene; formulation used as both glue and seam, being also applied to the outer fabric on the bonding region; and a sewn sample using zig zag stitch with a polyester thread as reference of traditional bonding method. The material was applied to both sides of the substrate using a syringe and a custom 3D printed nozzle, ensuring controlled seam width of 10 mm and maximal thickness of 1 mm along the samples. Schematic representations of all bonding methods are provided. The jointless neoprene has high failure stress of 3.25 MPa and strain at failure 235%. Performance of neoprene joints varied depending on the bonding method. The adhesive formulation as glue yielded a joint resisting 0.57, and strain of 80%.

Using the formulation as both glue and seam improves the joint (~0.94 MPa stress and ~106% strain) and yields comparable results to the sewn sample (~0.84 MPa stress and ~114% strain).

Additional advantages of the dual-network formulation are represented in **Figure 5c** by photos of both seams shortly before they start to fail: while the sewn sample quickly opens a large gap between the neoprene pieces, the punctures where the threads pass over are points of failure and opening – detrimental in cases where resistance to water penetration is necessary, for example. The dual-network adhesive sample, on the other hand, provides sealing up to large deformations and a more homogeneous deformation due to the outer seam shape, which creates a small gradient effect - thicker and stiffer at the center of the joint between the neoprene pieces, and thinner moving away from the seam and integrating to the substrate. Moreover, there is no delamination from the fabric, since the adhesive allows for penetration into the fabric prior to curing, as depicted in Figure 1a. Finally, **Figure 5d** depicts the Stress x Strain response of the neoprene fabric compared to the bonded samples. Using the 5 mol% formulation as a glue, the mechanical response matches that of the original material, as indicated by the (*), demonstrating a mechanically transparent seam.

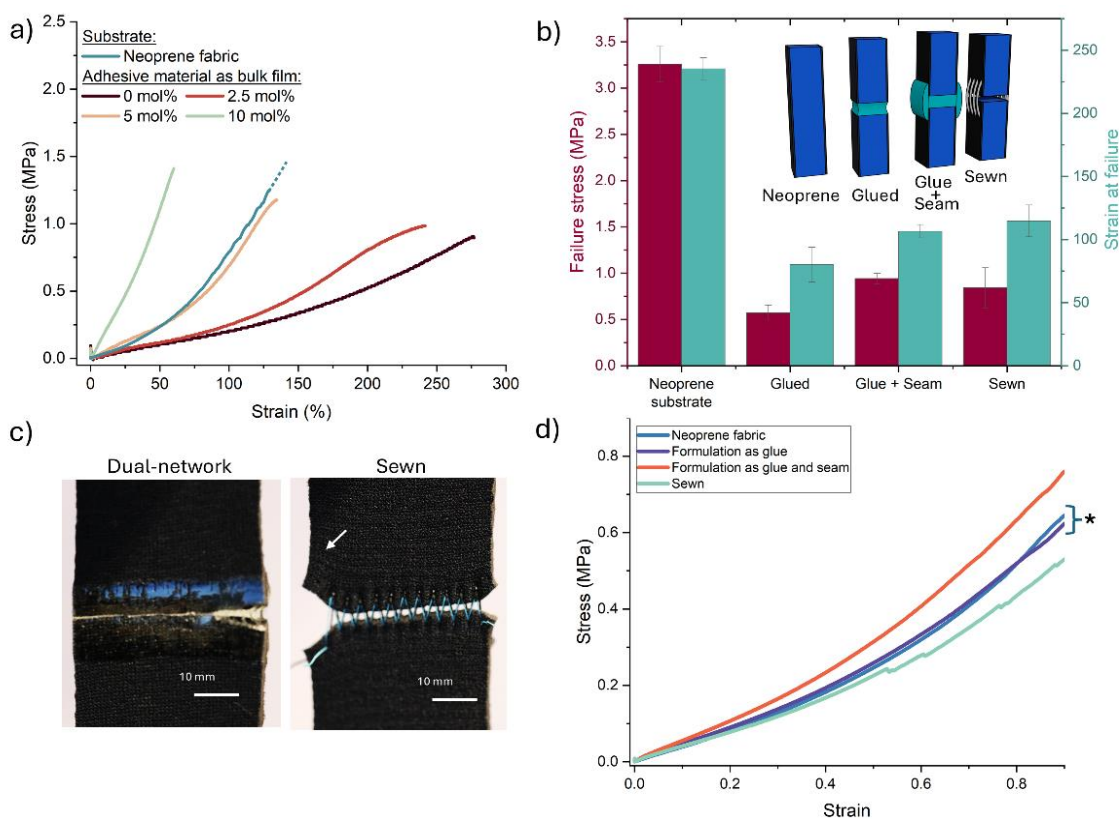

*Figure 5: Tensile test results of neoprene substrate attached under various bonding methods. a) Stress x strain curve for the formulations as bulk films exposed to UV for 15 min, compared to the pristine neoprene fabric substrate. b) Failure stress and strain at failure for the following configurations: (1) single piece of pure neoprene substrate without any adhesive (control), two pieces of neoprene: (2) bonded at the center using the adhesive as glue, (3) bonded with both glued center and external seam application of the adhesive (glue + seam), and (4) sewn together, representing current conventional attachment methods (sewn). The results highlight*

*the effectiveness of our acrylate-based adhesive in providing a strong, stretchable bond comparable to traditional sewing techniques. c) Photos of the joints glue + seam and sewn, shortly before they start to fail under tensile strain. White arrow indicates point of failure initiation caused by sewing puncture. d) Stress x strain curves for the bonded materials; curved indicated by the (\*) highlight that the glue material is capable of attaching two pieces of neoprene while maintaining mechanical response of the original material, creating a mechanically transparent seam.*

## 3. Conclusions

We combined design strategies of a dual network and high-order structuring to create a tunable adhesive material to match the properties of soft, stretchable substrates. The results presented in this study indicate a combined effect of (1) self-assembly of PMMA-b-PnBA-b-PMMA triblock copolymer, where PMMA end-blocks form microdomains and act as physical crosslinks, and (2) an *in situ* polymerized PnBA network where varying crosslinker content provide a range of strength and stretchability to match different substrates. In this dual network, the chemically crosslinked PnBA network adds resistance to failure, complementing the triblock copolymer which is self assembles into stiffer microdomains creating a physically crosslinked network, resulting in a strong, stretchable adhesive The *in situ* polymerization allows for effective integration of the PnBA secondary network, and dismisses the use of solvents, offering a more sustainable adhesive solution.

We demonstrated the application of the dual network as an adhesive for neoprene fabric by selecting the formulation with 5 mol% crosslinker, which, when applied as a glue between materials, matched the adherends' mechanical response, creating a mechanically transparent seam. When used as both glue and seam, the material showed comparable failure stress and strain to standard sewing, with the additional benefit of avoiding stress concentration points and providing barrier properties.

A more detailed study to understand quantitative contributions of both effects - self-assembly mechanisms, and composition contribution - is required for a more rational design of this novel system for adhesives and other nanostructured applications. Understanding those mechanistic aspects should clarify the contribution of the crosslinking density of the secondary network and assist in the molecular design of the material for more precise material development. Moreover, the overall seam integrity and performance can be improved by applying the adhesive in a controlled manner, for instance creating a gradient of physical and chemical properties, with varying crosslink density and stiffness, through different application designs or thickness gradients.

The unique properties and solvent-free, simple preparation of this adhesive make it a sustainable alternative for various complex substrates. Its ability to create a mechanically transparent seam ensures that the adhesive integrate seamlessly into stretchable materials, avoiding stress concentrations and preserving original mechanical properties. Our material offers a greener solution for current methods used in garments, sportswear, outdoor equipment, soft robotics, civil engineering, and other fields where stretchable and soft adhesives are required.

## 4. Experimental section

*Materials:* ABA Triblock copolymers consisting of poly(methyl methacrylate) (PMMA) end blocks and poly(n-butyl acrylate) (PnBA) mid-block (Kurarity, provided by Kuraray Co., Ltd) were used as received. Butyl acrylate (BA) monomer and ethylene glycol dimethacrylate (EGDMA) crosslinker were purchased from Sigma Aldrich. The photoactive photoinitiator Irgacure 819 was acquired from Ciba-Giegy Specialty Chemical Divisions. BA and EGDMA were passed through basic alumina (active, 32-63 µm, Sorbtech) to remove the inhibitor before polymerization. The photoinitiator was used as received. A neoprene fabric substrate (Sewswank Store) with 3 mm thickness was used for adhesion tests.

*Triblock copolymer selection:* The triblock copolymers LA2250, LA2140 and LA2230 were dissolved in 2-ethyl-1-hexanol (Sigma Aldrich) in 20 wt%, concentration thoroughly dried for 10 days at room temperature and tested using DMA (Figure S6). LA2250 Kurarity was selected for the adhesive formulations due to its higher storage modulus and nominal higher PMMA content. Gel permeation chromatography (GPC) was used to determine the average molecular weight of the selected triblock copolymer (Figure 2).

*Sample fabrication:* A mixture of triblock pellets and butyl acrylate (BA) monomer (M1) was prepared at least 24 hour prior to use, with an initial composition of 40 wt% triblock polymer and 60 wt% BA. Higher concentrations of triblock in BA were not explored due to substantial increase in solution viscosity, which adversely affected processability. The crosslinker and initiator were subsequently added to M1 and mixed homogeneously using a speed mixer (DAC-330-100 Pro). The amount of crosslinker added was calculated as a fraction of the BA molar content (2.5 mol%, 5 mol%, 10 mol%) while the amount of initiator was calculated based on the number of vinyl groups in each sample (1 mol% or 2 mol%). For example, a sample with 2 mol% initiator contains (0.02 x (mol BA + 2x mol EGDMA)) mol initiator.

To verify the effects of initiator content, samples without crosslinker (0 mol%) were tested with either 1 mol% or 2 mol% initiator. To assess the effects of crosslinker, all crosslinked samples were tested with 2 mol% initiator. Table 1 indicates the weight fraction of each component in the final adhesive formulation.

*Table 1: Weight fractions of the components in the final adhesive formulation, showing the distribution of triblock copolymer, butyl acrylate (BA), crosslinker (EGDMA), and initiator (Irgacure 819) for various sample compositions.*

| Samples | Weight fraction in final formulation | | | |
| --- | --- | --- | --- | --- |
| | Triblock | BA | EGDMA | Initiator |
| Initiator 1 mol% | 0.39 | 0.59 | 0 | 0.02 |
| Initiator 2 mol%, Crosslinker 0 mol% | 0.38 | 0.58 | 0 | 0.04 |
| Crosslinker 2.5 mol% | 0.37 | 0.55 | 0.04 | 0.04 |
| Crosslinker 5 mol% | 0.35 | 0.53 | 0.08 | 0.04 |
| Crosslinker 10 mol% | 0.32 | 0.48 | 0.15 | 0.04 |

The formulation was cast into molds with a polytetrafluoroethylene (PTFE, Teflon) bottom and leveled using a glass slide to achieve a uniform thickness and smooth surface for preparing "bulk samples". For the adhesion experiments, the material was applied to the adherend substrates using a syringe and a 3D printed polylactic acid (PLA) nozzle, ensuring controlled seam width (10 mm) and max thickness (1 mm) along 40 mm wide neoprene samples, applied on both sides. The materials were cured under UV-light for 5, 15 or 30 minutes using a near UV illumination system equipped with an Arc Mercury lamp 500 W (Newport #6285 = USHIO USH-508SA), with the intensity set to 7.5 to 9 mW/cm². For gravimetric measurements, one set of samples was placed in the oven at 40 °C overnight and weighed before and after the post-baking treatment.

*Characterization:*

*Gel permeation chromatography* (GPC) was conducted in an Agilent Technologies 12160 Infinity series system with two 5 μm mixed-D columns, a 5 μm guard column, a PL Gel 5 μm analytical mixed-D column, and a refractive index (RI) detector (HP1047A); tetrahydrofuran (THF) was used as eluent with a flow rate of 1.0 ml/min, PMMA standards were used for the calibration.

*Infrared spectra* (IR) were recorded using 64 scans, in a PerkinElmer Spectrum 100 FT-IR spectrometer in ATR mode using a diamond ATR element.

*Thermogravimetric analysis* (TGA) was performed using a TGA Q50 from TA Instruments. Samples were heated from room temperature to 250 °C at a heating rate of 2 °C/min under nitrogen. Thermogravimetric mass loss (TG) and mass loss derivative curve (DTG) were recorded as a function of time and temperature.

*Small-Angle X-ray scattering (SAXS)* experiments were performed on the cast "bulk films" of approximately 1 mm thickness using a Rigaku SmartLab X-ray Diffractometer with a CuK$\alpha$ source ($\lambda$=0.159 nm). Scattering vector range covered a range from q=0 to 0.3 using a beam of 3 kW. Data was collected in a HyPix3000 Hybrid Pixel Array detector.

*Transmission electron microscopy (TEM)* sample preparation was performed by cutting ultrathin sections (ca. 90 nm thick) with a glass knife on a Leica Ultracut UCT microtome under cryogenic conditions at -90 °C. Sections were collected on 400 mesh carbon coated copper grids, and the PMMA regions were subsequently stained using an aqueous solution of 2 wt% phosphotungstic acid (PTA) and 2 wt% benzyl alcohol at room temperature for 10 minutes.[34] Stained samples were imaged in a transmission electron microscope JEOL JEM-2200FS. TEM images were analyzed using a custom python code and OpenCV python library for denoising, thresholding and object selection. Scikit-learn[35] was used for nearest neighbor calculations.

*Dynamic Mechanical analysis (DMA)* of the cured adhesive formulations was conducted with a Discovery DMA 850 (TA Instruments) at room temperature. Rectangular samples with a thickness between 0.5 and 1 mm and width of 5 mm were placed between the tension clamps and storage moduli (E'), loss moduli (E''), and tan δ (= E''/E') were measured with varying frequency from 0.1 to 100 Hz at fixed strain 0.1% and preload force 0.01 N.

*Uniaxial tensile testing* was conducted using an Instron 68TM-5 with an advanced video extensometer (AVE 2) for precise digital strain measurement. Samples underwent controlled deformation at 305 mm/s until failure. Bulk samples were cut into rectangular samples with 10 mm width, thickness ranging from 0.6 to 1 mm, and measured using a gauge length of 50 mm. Neoprene fabric substrates (70% Neoprene, 30% Nylon) were purchased from Macro International. A control sample was prepared by sewing two pieces of neoprene using a zig-zag stitch with polyester thread in a Singer Confidence 7470 sewing machine.

**Supporting Information**

Supporting Information is available from the Wiley Online Library or from the author.

**Conflict of Interest**

The authors declare no conflict of interest.

**Acknowledgements**

The authors thank Sheico Group and CUMIRP for financial support of this work, and Dr. David Waldman for his assistance in facilitating industrial collaborations. The authors would like to thank Kuraray for providing the triblock copolymers used in this work. We would like to thank Prof. Ramaswamy Nagarajan and the support from the Center for Advanced Materials (CAM) and HEROES Initiative at the University of Massachusetts Lowell for the SAXS measurements. We thank Dr. Alexander Ribbe, Connor Witt, and Zaw Htet Lin for their technical support with TEM analysis and sample preparation. The authors would like to thank Dr. Hyemin Kim, Prof. Han-Yu Hsueh and Ting-Lun Chen for helpful discussions.

Facilities used during this research are maintained by the University of Massachusetts Amherst.

Received: ((will be filled in by the editorial staff))
Revised: ((will be filled in by the editorial staff))
Published online: ((will be filled in by the editorial staff))

Table of Contents ToC:

**Architected Dual-Network Solvent-free Adhesives for Stretchable Fabrics**

Gabriela Moreira Lana, Cornelia Meissner, Siddhant Iyer, Xin Hu, Perin Jhaveri, Skylar Tibbits, Alfred J. Crosby

We present a sustainable solvent-free, dual-network, acrylate-based adhesive that combines a triblock copolymer self-assembly with an in-situ polymerized second network. The physical and chemical crosslinking provide tunable strength and stretchability, making it ideal for high-performance bonding of soft, complex adherends like fabrics.

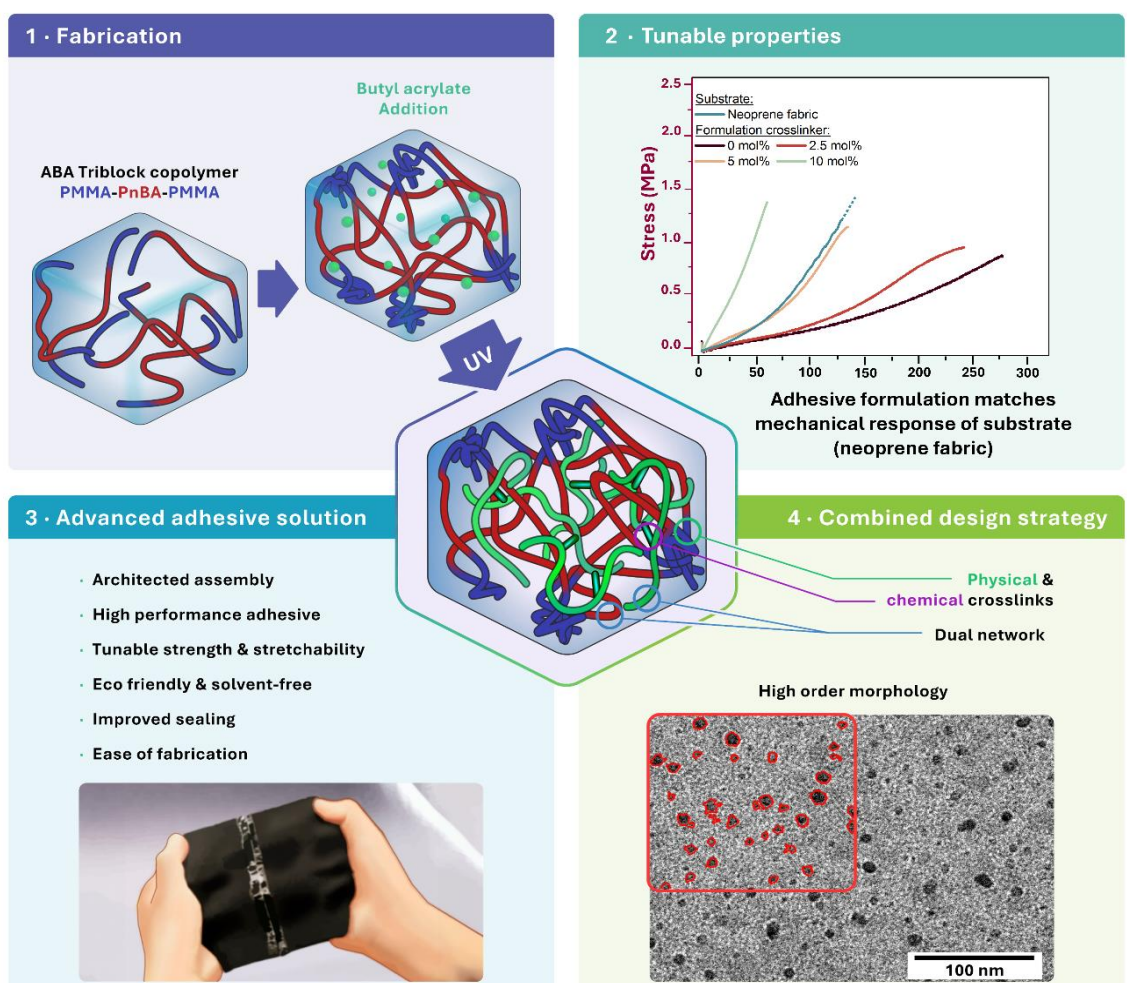

# Supporting Information

**Architected Dual-Network Solvent-free Adhesives for Stretchable Fabrics**

Gabriela Moreira Lana, Cornelia Meissner, Siddhant Iyer, Xin Hu, Perin Jhaveri, Skylar Tibbits, Alfred J. Crosby

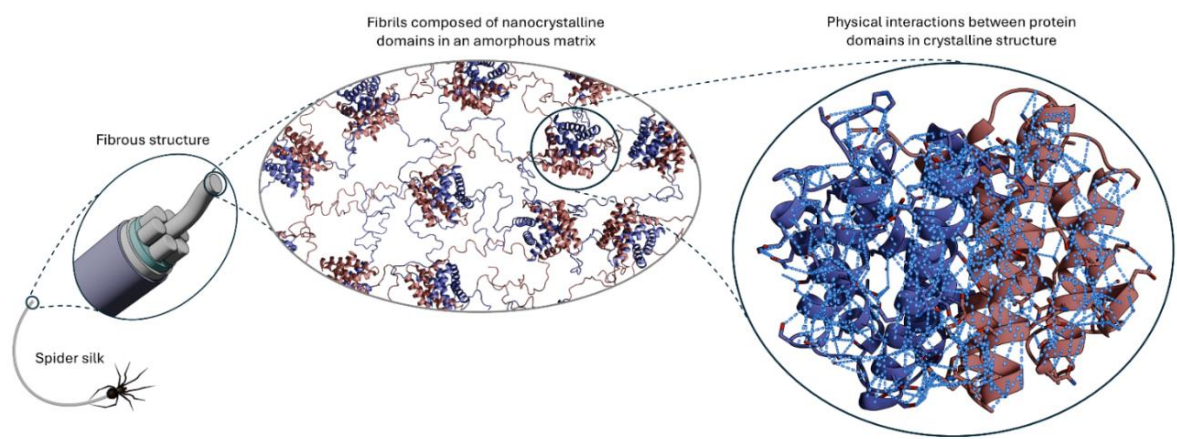

**Figure S1**: Schematic representation of natural inspiration for strong and flexible dual networks: spider silk hierarchical structure with stiff nanocrystalline β-sheet regions within a softer, amorphous matrix.

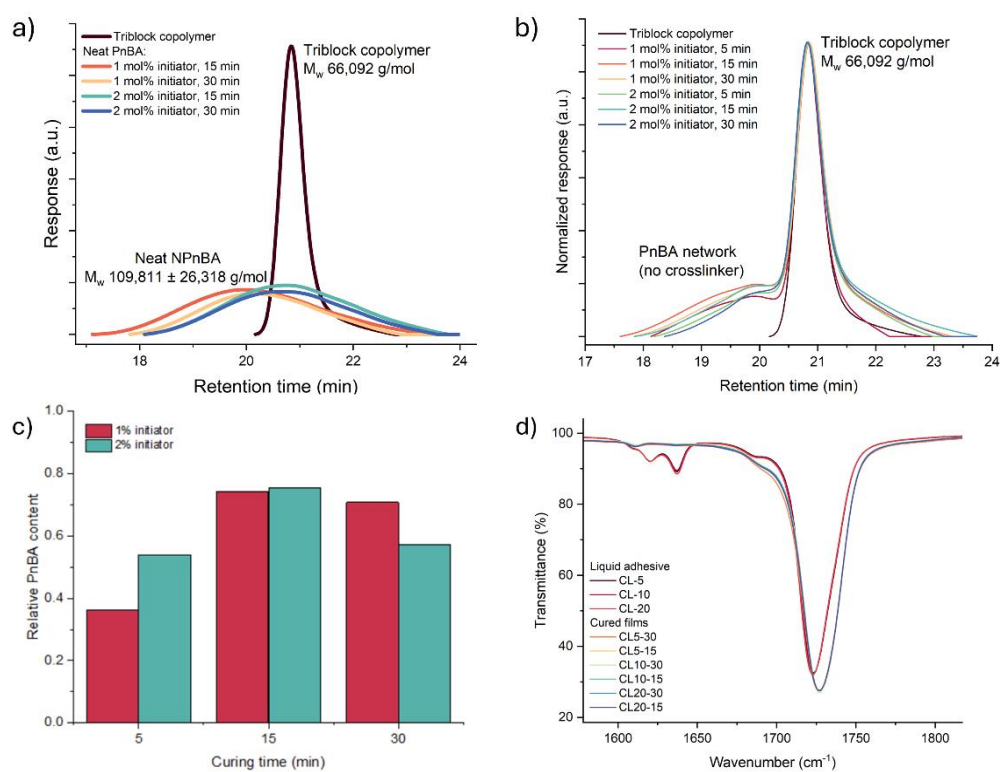

**Figure S2:** a) GPC profiles of triblock copolymer in comparison to neat PnBA polymerized with 1 and 2 mol% initiator, cured under UV light for 5, 15 and 30 min. B) GPC profiles of triblock in comparison to samples with 1 mol% or 2 mol% initiator and polymerization times

5, 15 and 30 min. C) Relative PnBA content of samples with 1 mol% or 2 mol% initiator, calculated from the area under the curves in Figure S2-b. Due to low yield for samples cured for 5 min, we proceeded with samples polymerized for 15 and 30 min. d) FTIR of liquid adhesive formulation and crosslinked samples. C=C bond region between 1600 and 1650 cm$^{-1}$ is present for the liquid phase, and absent after UV polymerization for 15 and 30 min, indicating complete reaction.

Table S1: Number-average (Mn) and weight-average (Mw) molecular weight and molecular weight distribution, determined by the polydispersity (PD) of the triblock copolymer sample and the neat poly n-butyl acrylate at different fabrication conditions, determined by GPC.

| Neat triblock copolymer PMMA-b-PnBA-b-PMAA ||||
| $M_n$ (g/mol) || $M_w$ (g/mol) | PD |
| 62,396 || 66,092 | 1.07 |
| Neat Poly-n-butyl acrylate |||||
| Initiator (mol%) | UV time (min) | $M_n$ (g/mol) | $M_w$ (g/mol) | PD |
| 1 | 5 | 51,862 | 85,211 | 1.643 |
| 1 | 15 | 78,262 | 158,080 | 2.0199 |
| 1 | 30 | 79,870 | 132,191 | 1.6551 |
| 2 | 5 | 56,623 | 92,939 | 1.7546 |
| 2 | 15 | 50,292 | 93,028 | 1.8498 |
| 2 | 30 | 54,631 | 97,414 | 1.7831 |

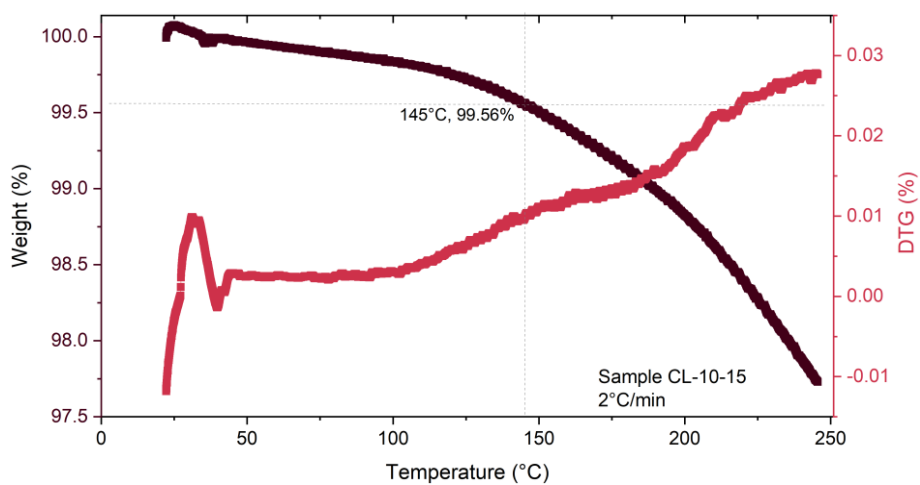

**Figure S3:** TG and DTG curves of sample with 5 mol% crosslinker, exposed to UV for 15 min. Weight loss at 145 °C (Boiling point of butyl acrylate) is 0.44 %.

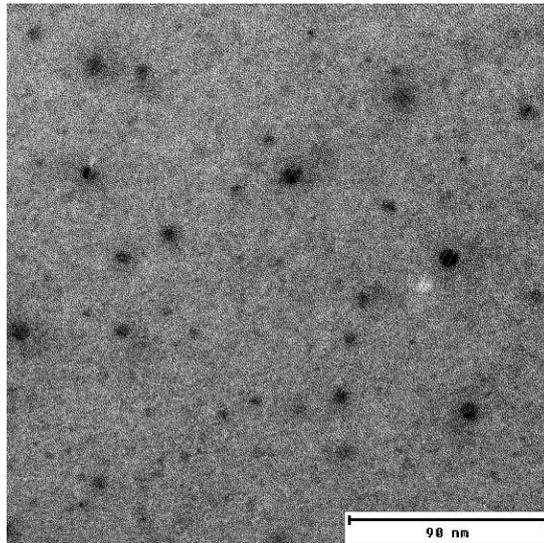

**Figure S4**: TEM image of neat triblock copolymer. Bulk sample prepared by dissolving 20 wt% polymer in 2-ethyl-1-hexanol and drying at room temperature for 10 days.

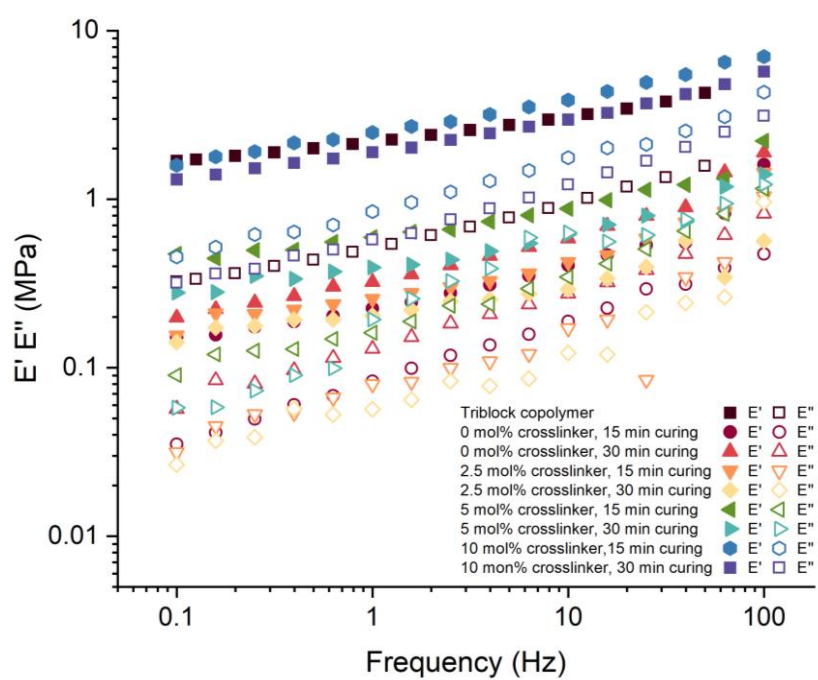

**Figure S5**: Mechanical characterization of films using DMA. Crosslinker content ranging from 0 mol% to 10 mol%, and UV polymerization times 15 min and 30 min.

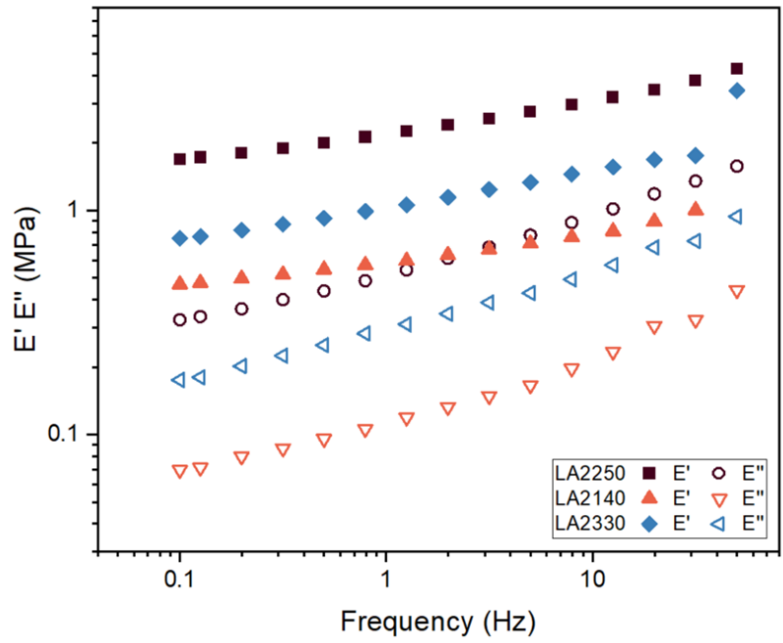

**Figure S6:** DMA results of triblock copolymer materials from Kuraray, storage (E') and loss modulus (E'') as function of frequency. LA2250 was selected for this work.